\documentclass[aps,prb,twocolumn,groupedaddress,showkeys,showpacs,superscriptaddress,floatfix]{revtex4-1}
\usepackage{epsfig}
\usepackage{lipsum}
\usepackage{multirow}
\usepackage{amsmath, amssymb,mathtools}
\usepackage{color}
\usepackage{graphicx}
\usepackage{dcolumn}
\usepackage{bm}
\usepackage{subfigure}
\usepackage[utf8]{inputenc}
\usepackage[T1]{fontenc}
\usepackage{tikz}

\graphicspath{{Figures/}}

\begin{document}

\title{Frustration driven Josephson phase dynamics}

\author{Claudio Guarcello}
\email{cguarcello@unisa.it}
\affiliation{Dipartimento di Fisica “E. R. Caianiello,” Università di Salerno, Via Giovanni Paolo II 132, I-84084 Fisciano (SA), Italy}
\affiliation{INFN, Sezione di Napoli Gruppo Collegato di Salerno, Complesso Universitario di Monte S. Angelo, I-80126 Napoli, Italy}

\author{Luca Chirolli}
\email{luca.chirolli@nano.cnr.it}
\affiliation{Department of Physics, University of California, Berkeley, California 94720, USA}
\affiliation{NEST, Istituto Nanoscienze-CNR and Scuola Normale Superiore, I-56127 Pisa, Italy}

\author{Maria Teresa Mercaldo}
\email{mmercaldo@unisa.it}
\affiliation{Dipartimento di Fisica “E. R. Caianiello,” Università di Salerno, Via Giovanni Paolo II 132, I-84084 Fisciano (SA), Italy}

\author{Francesco Giazotto}
\email{francesco.giazotto@nano.cnr.it}
\affiliation{NEST, Istituto Nanoscienze-CNR and Scuola Normale Superiore, I-56127 Pisa, Italy}

\author{Mario Cuoco}
\email{mario.cuoco@spin.cnr.it}
\affiliation{SPIN-CNR, c/o Università di Salerno, I-84084 Fisciano (SA), Italy}

\date{\today}

\begin{abstract}
The Josephson equations predict remarkable effects concerning the phase state of a superconducting junction with an oscillating current induced by a static voltage. Whether the paradigm can be twisted by yielding an oscillating voltage without making use of harmonic drives is a fundamentally relevant problem yet not fully settled. Here, we demonstrate that a dynamical regime with an oscillating phase evolution is a general hallmark of driven Josephson systems exhibiting sign competition in the Josephson couplings. We show that in frustrated Josephson systems an oscillating phase dynamics gets switched on by driving the changeover among different ground states, which can be induced by varying the parameters that set the phase state. Remarkably, the character of the transitions in the Josephson phase space allows different types of dynamics, with few or several harmonics. This result sets out a characteristic mark of any superconducting system with frustrated Josephson couplings and can be exploited to disentangle the complexity of the underlying phases. 
\end{abstract}

\maketitle

\section{Introduction} 

A Josephson junction (JJ) allows to couple the phase of coherent paired states 
in two weakly linked superconductors with experimentally accessible quantities, such as the flowing supercurrent and the voltage drop, i.e., the well-known Josephson relations \cite{Jos62,Jos74}. The voltage drop across the device sets out the rate at which the Josephson phase evolves in time; in fact, a direct conversion of a static dc-voltage into high-frequency electromagnetic oscillation of the Josephson current can be attained. 

Starting from the consolidated Josephson effects, a fundamental and different perspective points to whether the paradigm can be reversed by having, instead of a current, an oscillating voltage, or both current and voltage oscillating in time, without making use of harmonic drives. Such scenario poses also key questions, not yet fully settled, about the mechanisms or the Josephson setups that can be employed to achieve this type of dynamical regime. 
Here, we tackle this challenge and demonstrate that a dynamical regime with an oscillating phase evolution is indeed a general hallmark of Josephson driven systems that exhibit sign frustration in the Josephson couplings without externally applied current/voltage bias. In particular, we demonstrate the establishment of time-dependent coherent or incoherent phase dynamics in response to a linear in time adiabatic perturbation.

Superconducting systems with unconventional phase relations are quite ubiquitous in condensed matter.
A special role in this context is played by the so called $\pi$-phase shifts and $\pi$-pairing, i.e., antiphase relation between order parameters or equivalently the sign reversal of the effective Josephson coupling between Cooper pairs. This is at the heart of unconventional superconductivity, e.g., in cuprates \cite{Kir11,TsuKir00}, iron-based \cite{Gri20,Gri21} and oxide interface superconductors \cite{Sche15,Sin21}, superconductor-ferromagnet-superconductor junctions \cite{Feo10}, phase qubits~\cite{Gin16}, electrically or orbitally driven superconducting phases \cite{Mer20,Bou20,DeSim21,Mer21}, and multi-orbital non-centrosymmetric superconductors \cite{Fuk18,Fuk20,Sche15,Mer20}. 
However, when there is no simple phase ordering pattern that satisfies all Josephson couplings, the unsatisfied one is said to be frustrated.
Along this line, disentangling the complexity arising from superconducting phase frustration in the presence of $0$ and $\pi$-pairings is a demanding and non-trivial achievement~\cite{Gri20,Gri21,Tri21,Sin21}. 
The frustrated Josephson coupled systems composed of 0- and $\pi$- JJs were already investigated~\cite{Dia11,Dia14}, even considering frustrated multi-bands superconductors and in the case of arrays of JJs~\cite{Dia14,And19}, where the presence of both degenerate and non-degenerate ground states was also discussed~\cite{And19}.

To this aim, we show that in frustrated Josephson systems an oscillating phase dynamics gets switched on by driving the changeover among ground states in the phase space and can be guided by varying the parameters that set the phase state, e.g., the Josephson couplings.
A remarkable fingerprint of these oscillating-phase regimes is that they can be toggled from coherent to incoherent in the time dependence by selecting the type of transition in the Josephson phase space. These marks can be exploited to single out the presence and the character of superconducting phase frustration in intrinsic or engineered superconducting systems~\cite{Kon02,Wei06} as well as the nature of the resulting ground state. The investigated dynamical behavior is also predicted to occur for transitions involving degenerate ground states, as in the so called $\varphi$-JJ~\cite{Buz03,Sic12}. Finally, we note that the phenomenon described in this work bears a certain similarity to the synchronization phenomenon that occurs in arrays of interacting JJs.~\cite{Lik86,Wie96}.

\begin{figure}[t!!]
\includegraphics[width=\columnwidth]{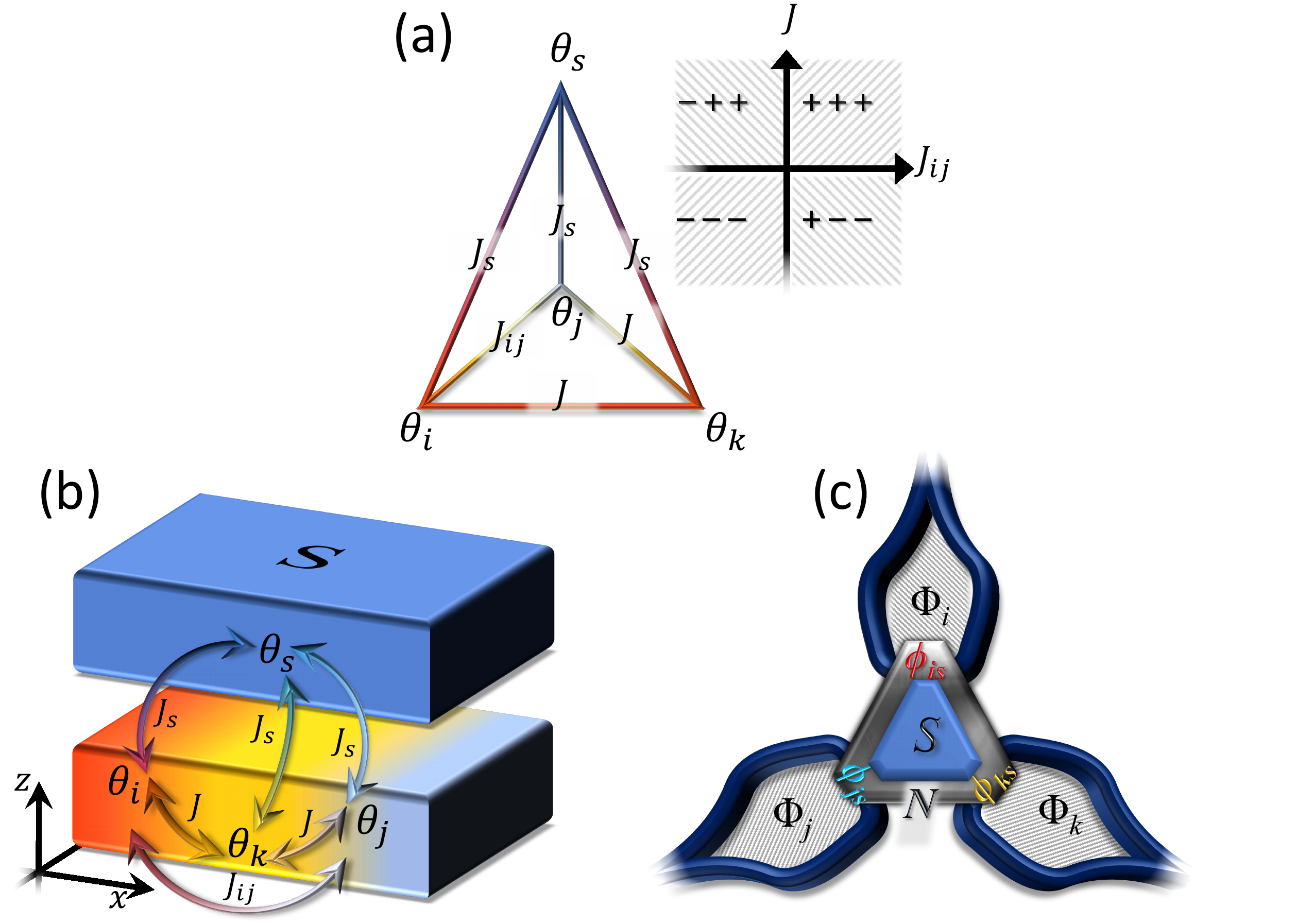}
\caption{(a) Schematic representation of Josephson phases assuming four phase degrees of freedom and sign-competing Josephson couplings, i.e $J_s$, $J_{ij}$, $J$. (b) Sketch of the JJ made of a three-band superconductor with $\pi$-pairing and an \emph{s}-wave single-band superconductor. The interband, $J$ and $J_{ij}$, and interjunction, $J_s$, couplings are highlighted. (c) Equivalent circuit of the multiband JJ. The fluxes $\Phi_z$ are used to establish the $\pi$-pairings.}\label{Fig01}
\end{figure}

The paper is organized as follows. In Sec.~\ref{Sec01}, we describe the Josephson system and the ground states. In Sec.~\ref{Sec02}, we introduce the phase dynamics triggered in the case of a few specific transitions and the frequency response. In Sec.~\ref{Sec03}, the conclusions are drawn.

\section{Model}
\label{Sec01}

A variety of Josephson-based systems characterized by phase competition has been reported in literature~\cite{Dia11,Lin12,Wes13,Hua14,Boj14,Lin14,Tan15,Yer17} mostly focusing on two competing Josephson channels. Here, we consider an effective model with three coupled Josephson channels having 0- or $\pi$-character [Fig.~\ref{Fig01}(a)]. This scenario can be directly implemented by considering a junction made of an \emph{s}-wave superconductor interfaced to a multiband superconductor~\cite{Tan15,Yer17} [Fig.~\ref{Fig01}(b)] or, equivalently, a superconducting circuit [Fig.~\ref{Fig01}(c)] designed by connecting, via normal channels, a central superconducting island to three superconducting electrodes, which are reciprocally coupled, and whose phases can be modulated by magnetic fluxes.
We consider a multicomponent junction based on three superconducting Josephson channels, $J_{zs}$
with $z=i,j,k$, and $\phi_{zs}=\theta_z-\theta_s$ indicating the relative phases across the junction. $\theta_z$  and $\theta_s$ stand for the phases of the three-band and the \emph{s}-wave superconductor, respectively [e.g., Fig.~\ref{Fig01}(b)]. The relative phases between $\theta_i$, $\theta_j$, and $\theta_k$ are set out by the internal degrees of freedom of the superconductor, which can be due to non-conventional pairing glues, electronic reconstruction, or externally driven sources of symmetry breaking. 
The interband Josephson couplings, i.e., established between different order parameters of the three-band superconductor, can be positive or negative, the latter in the case of a $\pi$-pairing. The occurrence of these $\pi$-couplings can lead to frustrated configuration. Frustration arises here from the impossibility of having all interactions favorable.

\begin{figure}[t!!]
\includegraphics[width=\columnwidth]{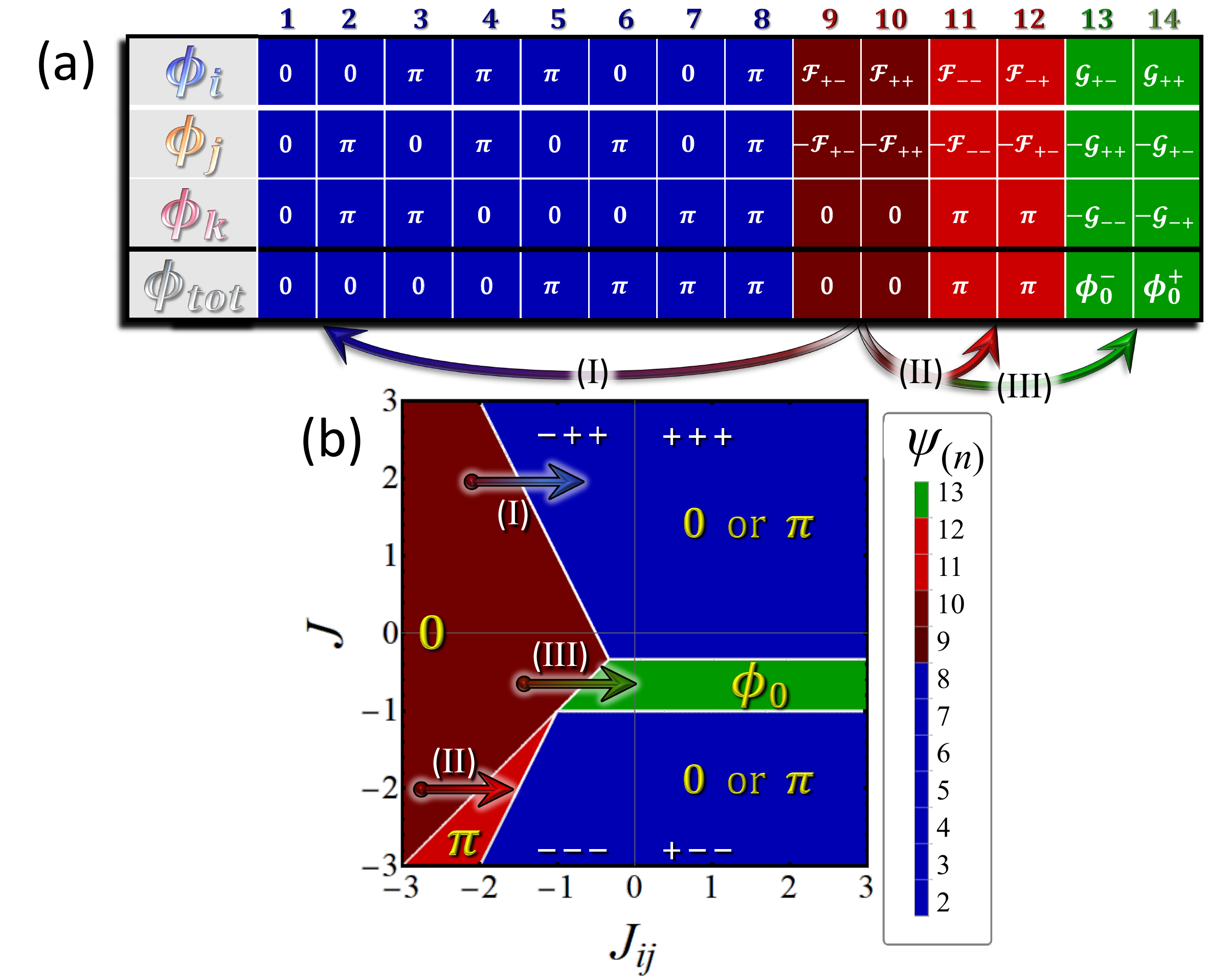}
\caption{(a) GSs $\psi_{(n)}=(\phi_i,\phi_j,\phi_k)_{(n)}$, with $n=1\dots14$, and a total phase value $\phi_{tot}$. (b) Phase diagram of the lowest-energy GS as a function of $J$ and $J_{ij}$. The arrows highlight the phase transitions discussed in Fig.~\ref{Fig03} and labeled with (I), (II), and (III). The total phase values $\phi_{tot}=\{0,\pi,\text{ or }\phi_0\}$ in the GSs are also indicated.}
\label{Fig02}
\end{figure}

In the absence of magnetic field and bias current, the total Josephson energy is
\begin{eqnarray}\label{EqA01}
E=&-&\sum_{z=i,j,k}J_{zs}\cos\phi_{zs}-J_{ij}\cos\left ( \phi_{is}-\phi_{js} \right )\\\nonumber
&-&J_{ik}\cos\left ( \phi_{is}-\phi_{ks} \right )-J_{jk}\cos\left ( \phi_{js}-\phi_{ks} \right ).\nonumber
\end{eqnarray}

\begin{figure*}[t!!]
\includegraphics[width=2\columnwidth]{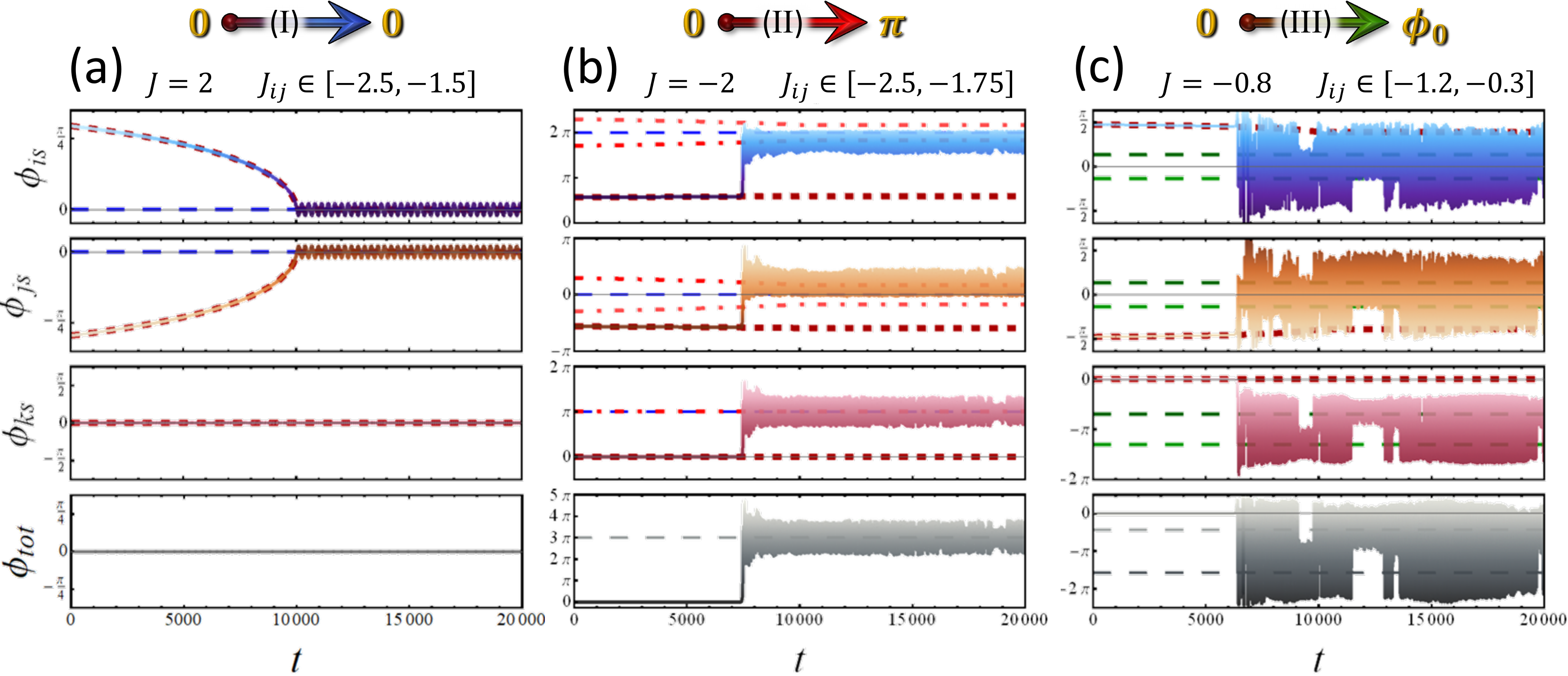}
\caption{Time dependent phase evolution for three different physical cases: (a) a $0\to0$ transition driven by setting $J=2$ and ranging $J_{ij}(t)\in[-2.5,-1.5]$, (b) a $0\to\pi$ transition driven by setting $J=-2$ and ranging $J_{ij}(t)\in[-2.5,-1.75]$, and (c) a $0\to\phi_0$ transition driven by setting $J=-0.8$ and ranging $J_{ij}(t)\in[-1.2,-0.3]$. The blue, red, and green dashed lines indicate the analytical solutions listed in Fig.~\ref{Fig02}(a) around which the phases evolve.}
\label{Fig03}
\end{figure*}

The vector $\psi_{(n)}=(\phi_i,\phi_j,\phi_k)_{(n)}$ defines the ground state (GS) configurations and can be obtained by minimizing the total energy with respect to $\theta_s$, $\theta_i$, and $\theta_j$. In particular, assuming equal interjunction contributions, i.e., $J_{is}=J_{js}=J_{ks}=J_{s}>0$, and that two of the three interband coupling coincide, i.e, $J_{ik}=J_{jk}=J$, one can get analytical expressions for the $\psi_{(n)}$, with $n=1\dots14$. 
These solutions can be in turn grouped in three classes, as reported in the table in Fig.~\ref{Fig02}(a). First, the system admits solutions that are uniquely given by combinations of $0$ and $\pm\pi$ [see the blue columns in Fig.~\ref{Fig02}(a) with $n$ from $1$ to $8$] that we refer as trivial since they correspond to standard time-reversal symmetric Josephson phase values. 
Then, two classes of non-trivial solutions emerge with the Josephson phases being not pinned to $0$ or $\pi$, thus yielding a configuration that breaks time-reversal symmetry. One class of configurations is given by $\phi_i=-\phi_j$, while $\phi_k = 0 \text{ or } \pi$ [see the red columns in Fig.~\ref{Fig02}(a) with $n$ from $9$ to $12$], given by 
\begin{equation}\label{EqA04}
(\phi_i,\phi_j,\phi_k)_{(n)}=(\mathcal{F}_{\sigma,\chi},-\mathcal{F}_{\sigma,\chi},0\text{ or }\pi)
\end{equation}
where $\sigma=\pm1$, $\chi=\pm1$, and $\mathcal{F}_{\sigma,\chi}=\arctan\left [ f_\sigma,\widetilde{f}_{\sigma,\chi} \right ]$~\footnote{$\arctan[x,y]$ gives the arctangent of $y/x$, taking into account which quadrant the point $(x,y)$ is in.}, with
\begin{equation}\nonumber
f_\sigma=-\frac{J_s+\sigma J}{J_{ij}}\qquad
\widetilde{f}_{\sigma,\chi}= \frac{\chi}{J_{ij}}\sqrt{\left (2J_{ij} \right )^2- \left (J+\sigma J_s \right )^2 }.
\end{equation}
Another class has all three phases with values different from $0$ or $\pi$ [see green columns in Fig.~\ref{Fig02}(a) with $n=13$ and $14$], which can be written as 
\begin{equation}\label{EqA05}
(\phi_i,\phi_j,\phi_k)_{(n)}=(\mathcal{G}_{\sigma,\chi},-\mathcal{G}_{\sigma,-\chi},-\mathcal{G}_{-\sigma,\chi})
\end{equation}
where $\sigma=\pm1$, $\chi=\pm1$, and $\mathcal{G}_{\sigma,\chi}=\arctan\left [ g_\sigma,\widetilde{g}_{\chi} \right ]$, with
\begin{equation}\nonumber
g_\sigma\!=\!-\sigma\frac{3J^2-J_s^2}{JJ_s}\qquad
\widetilde{g}_{\chi}\!=\!\frac{\chi}{J J_s}\sqrt{\left ( 2JJ_s \right )^2-\left ( 3J^2-J_s^2 \right )^2}.
\end{equation}

The knowledge of the explicit expression of the solutions allows us to have a high degree of control of the possible transitions in the phase diagram as well as of the corresponding dynamics.

\section{Results.}
\label{Sec02}

The phase space (PS) in Fig.~\ref{Fig02}(b) is constructed by evaluating the lowest energy solution among all $\psi_{(n)}$ versus the Josephson couplings (see Appendix~\ref{AppA}). 
The achieved PS can be divided in different areas, bounded by sharp white-marked edges, in which the total phase $\phi_{tot}=\phi_i+\phi_j+\phi_k$ takes specific values: $0$ in the dark-red and blue regions, $\pi$ in the light-red and blue regions, and $\phi_0$ in the range $(0-\pi)$ in the green region.

For the full dynamical description of the system, we employ the equations of motion for the gauge-invariant phase differences, $\phi_{is}(t),\phi_{js}(t),$ and $\phi_{ks}(t)$~\footnote{In order to further clarify the notation, we point out that for the GSs configurations we use $\phi _i,\phi _j,$ and $\phi _k$, while the time-dependent solutions of the differential equations are indicated with $\phi _{is}(t),\phi _{js}(t),$ and $\phi _{ks}(t)$}. 
The corresponding solutions can be derived from a Lagrangian approach along the line of the two-channel model presented in Ref.~\onlinecite{Lin12} (see Appendix~\ref{AppB} for more details).
In particular, we consider a short junction and the adiabatic change of coupling constants $J$ and $J_{ij}$, for driving a transition among different GSs across a phase boundary of the PS in Fig.~\ref{Fig02}(b).

\begin{figure*}[t!!]
\includegraphics[width=2\columnwidth]{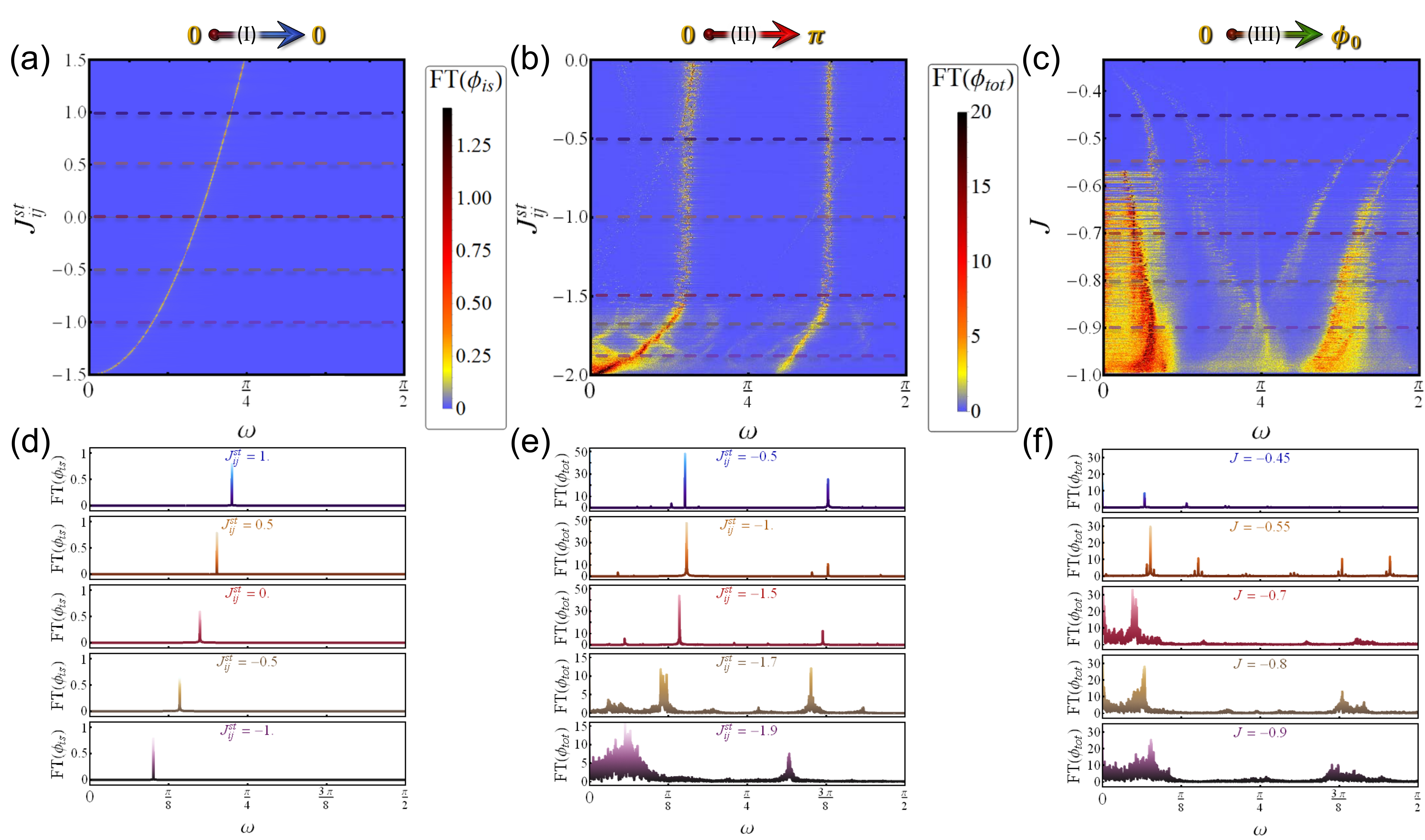}
\caption{Fourier transforms (FT) as a function of the coupling values in the three cases highlighted in Fig.~\ref{Fig02}(b): (a) FT of $\phi_i(t)$ as a function of $J_{ij}^{st}\in[-1.5,1.5]$, with $J=2$ and $J_{ij}(t)\in[-2.5,J_{ij}^{st}]$ and (d) some selected profiles; (b) FT of $\phi_{tot}(t)$ as a function of $J_{ij}^{st}\in[-2,0]$, with $J=-2$ and $J_{ij}(t)\in[-2.5,J_{ij}^{st}]$ and (e) some selected profiles; (c) FT of $\phi_{tot}(t)$ as a function of $J\in[-1,-0.33]$, with $J_{ij}(t)\in[J-0.5,J+0.5]$ and (f) some selected profiles. The legend in panel (b) refers also to panel (c).}
\label{Fig04}
\end{figure*}

In Fig.~\ref{Fig03} we collect the phase dynamics for three representative cases. In particular, as initial condition we choose the non-trivial GS $\psi_{(9)}$, with $\phi_{tot}=0$, which allows to drive a transition into all other configurations. This is done by setting three different $(J,J_{ij})$ trajectories. Then, by keeping constant the $J$ value, we adiabatically increase $J_{ij}(t)$, with a linear-in-time dependence, up to reach a specific value, $J_{ij}^{st}$, which is thereafter maintained fixed. 
The selected trajectories are highlighted by three arrows, labeled with (I), (II), and (III), in Fig.~\ref{Fig02}(b). They schematically depict how the $J_{ij}(t)$ are driven in order to induce the phase transitions shown in Fig.~\ref{Fig03}(a), (b), and (c), respectively [see Appendix~\ref{AppB} for a clear illustration of the $J_{ij}(t)$ drives]. 

In Fig.~\ref{Fig03}, dark-red-dashed lines mark the non-trivial GS $\psi_{(9)}$, while the blue-dashed lines identify the $0$ or $\pi$ trivial solutions, i.e., with $\phi_{tot}=0$ or $\pi$, the light-red-dot-dashed lines the non-trivial $\pi$-solutions, and the green-dashed lines the $\phi_0$-solutions. We observe that, in all cases shown in Fig.~\ref{Fig03}, initially the phases steadily follow the Josephson phase value of the ground state, i.e., the curves superimpose to the dark-red-dashed lines representing the $\psi_{(9)}$ GS. Then, approaching values of the Josephson couplings that correspond to the domain boundary in the PS, the phase evolution exhibits a dramatic change in the time dependence, with a behavior that is related to the character of the transition.

Figure~\ref{Fig03}(a), obtained for $J=2$ and ranging $J_{ij}(t)\in[-2.5,-1.5]$, demonstrates that even for a transition that conserves the global $\phi_{tot}$ value and occurs smoothly, one observes the appearance of a clear oscillating behavior in the $\phi_{is}(t)$ and $\phi_{js}(t)$ phases. These oscillations are triggered as the phases, initially matching the GS $\psi_{(9)}$, reach the trivial $0$ solution, specifically, the $\psi_{(1)}$ state having $\phi_i=\phi_j=\phi_k=0$. Interestingly, we observe that the oscillatory behaviors of $\phi_{is}(t)$ and $\phi_{js}(t)$ are equal but opposite in sign, so that the total phase $\phi_{tot}$ remains zero during the whole evolution. 

In Fig.~\ref{Fig03}(b) we illustrate the phase dynamics associated to a $0\to\pi$ transition in the PS, which can be obtained by setting $J=-2$ and varying $J_{ij}(t)$ in the range $[-2.5,-1.75]$. In this case, at a given time we clearly observe that the phase evolution exhibits a jump. After this steep variation, the phases oscillate around a non-trivial solution with a $\pi$ total phase, indicated by the light-red dot-dashed lines. Interestingly, also the total phase $\phi_{tot}$ undergoes a $\pi$-jump (all the $2\pi$ replica are equivalent), after which it starts to oscillate around a $\pi$-average value. 

Finally, Fig.~\ref{Fig03}(c) demonstrates that the time dynamics changes again by inducing a $0\to\phi_0$ transition in the PS, which can be achieved by choosing, for instance, $J=-0.8$ and ranging $J_{ij}(t)\in[-1.2,-0.3]$. Also in this case, the phases have a discontinuous time evolution. However, the state of the system thereafter oscillates between two distinct GSs, $\psi_{(n)}$ with $n=13$ and $14$. Interestingly, as the $0\to\phi_0$ transition occurs, the total phase $\phi_{tot}$ follows a similar evolution, starting to oscillate around the two predicted values, $\phi_0^\pm$ (see Appendix~\ref{AppB} for the full expressions of these quantities), which are indicated by gray dashed lines in the bottom panel of Fig.~\ref{Fig03}(c).

The character of the dynamical response can be deepened by investigating it in the frequency domain. In Fig.~\ref{Fig04}, we show the Fourier transforms (FT) of the phase signal after a transition occurs as a function of the Josephson couplings focusing on the three situations highlighted in Fig.~\ref{Fig02}(b), that is for the $0\to0$ (a), $0\to\pi$ (b), and $0\to\phi_0$ (c) transitions. 
In the bottom panels of Fig.~\ref{Fig04}, we include a few selected FT profiles traced in correspondence of the coupling values marked with the horizontal dashed lines in the counter plots shown in the top panels. 
In particular, in Fig.~\ref{Fig04}(a) we show the FT of $\phi_i(t)$ as a function of the steady value $J_{ij}^{st}$ taken by the time-dependent drive, i.e., we vary $J_{ij}^{st}\in[-1.5,1.5]$, keeping fixed $J=2$ and linearly ranging $J_{ij}(t)\in[-2.5,J_{ij}^{st}]$. For this trajectory, while the total phase steadily takes a zero amplitude, 
$\phi_i(t)$ exhibits an oscillating behavior in response to the $J_{ij}$-drive: in fact, Fig.~\ref{Fig04}(a) unveils a highly coherent response 
(for clarity, in Appendix~\ref{AppB} we show a few selected FT profiles).

The spectral profile is instead completely different as a transition $0\to\pi$ is considered. In Fig.~\ref{Fig04}(b) we report the Fourier spectra of $\phi_{tot}(t)$ by ranging $J_{ij}^{st}\in[-2,0]$, while taking $J=-2$ and $J_{ij}(t)\in[-2.5,J_{ij}^{st}]$. According to the value assumed by $J_{ij}^{st}$, two characteristic behaviors emerge. In fact, we find that for $J_{ij}^{st}\in[-2,-1.5]$ the frequency response of the system is significantly incoherent with several harmonics contributing to the dynamics. Conversely, for $J_{ij}^{st}\gtrsim-1.5$ the frequency spectrum is coherent, being composed by two sharp peaks. These behaviors reflect the steady states realized in the two cases. In fact, for $J_{ij}^{st}\in[-2,-1.5]$ the phases after the transition fluctuate around a non-trivial $\pi$-like configuration. On the other hand, when $J_{ij}^{st}\gtrsim-1.5$ after the transition the phases first temporarily linger on the non-trivial $\pi$-state, to then settle in a trivial $\pi$ state, whose ``position'' in the PS no longer depends on the coupling parameters. In the latter case, the total phase, $\phi_{tot}$, shows the coherent behavior as reported in Fig.~\ref{Fig04}(b).

\begin{figure}[t!!]
\includegraphics[width=0.9\columnwidth]{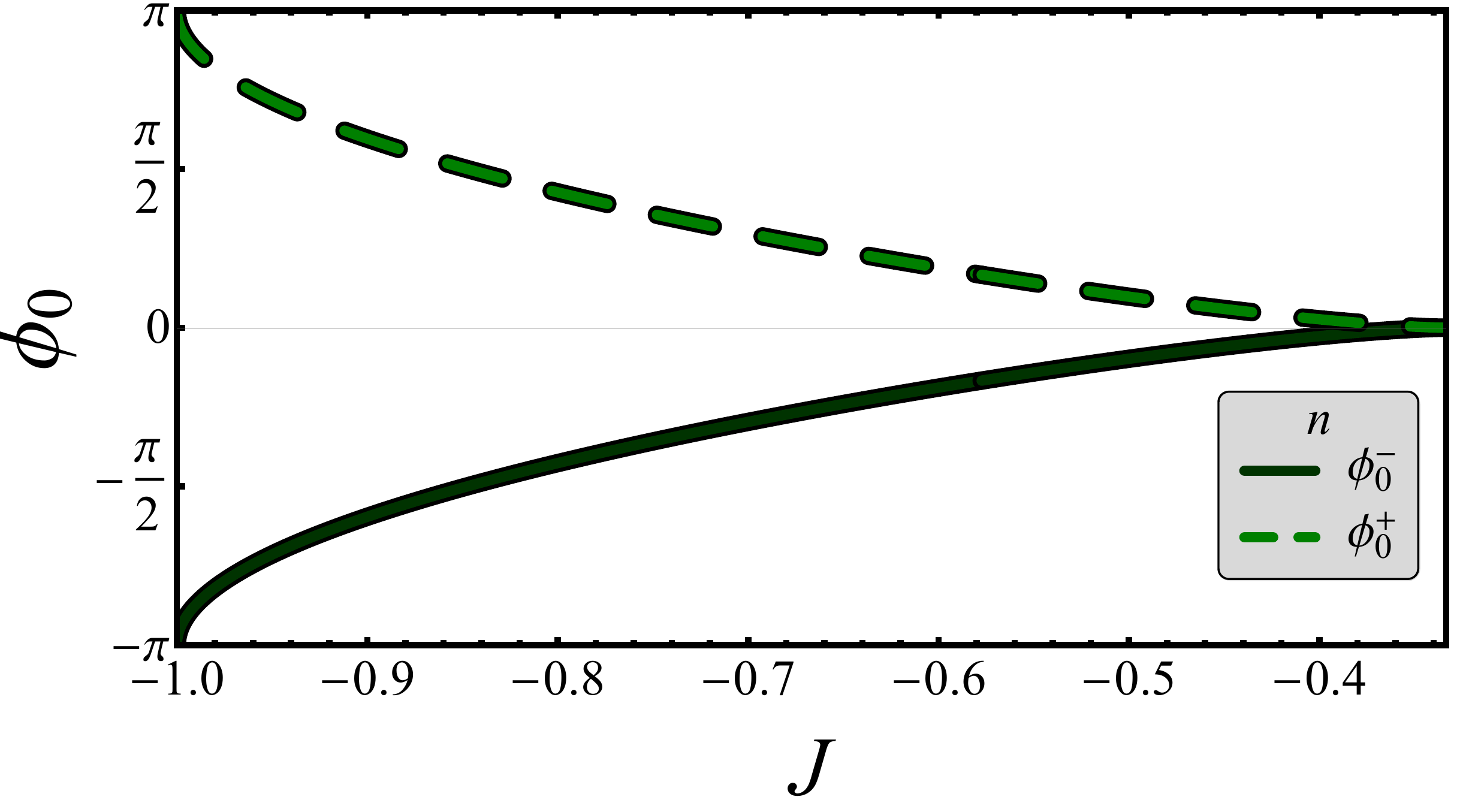}
\caption{Dependence of $\phi_0^{\pm}$ on the $J$ coupling value.}
\label{Fig05}
\end{figure}

Finally, we demonstrate in Fig.~\ref{Fig04}(c) how the frequency response gets modified in the case of a $0\to\phi_0$ transition. Here, we choose to explore the FT of $\phi_{tot}(t)$ by changing $J\in[-1,-0.33]$ and assuming a drive of the form $J_{ij}(t)\in[J-0.5,J+0.5]$. We observe that, also for this trajectory, the system can evolve in two different ways. For $J\in(-1,-0.6)$ the FT appears highly incoherent, with broad spectra composed by multi-peaked structures; conversely, for $J\in(-0.6,-1/3)$ the FT is characterized by few sharp peaks, whose positions shift towards zero as $J\to-1/3$, according to the fact that $\phi_0^{\pm}\to0$ in this case (see Fig.~\ref{Fig05}). 

We point out that these $\phi_0$ configurations essentially realizes a $\varphi$-type Josephson state~\cite{Buz03}. 
In this context, our study brings two general observations. Firstly, $\varphi$-degenerate ground states can be obtained without exploiting second-harmonic Josephson couplings as in setups using ferromagnetic layers~\cite{Sic12,Sto18} or ad-hoc geometries~\cite{Lip14,Gol15}. Second, any $\varphi$-junction that is driven from non-degenerate to degenerate phase configurations is expected to exhibit incoherent phase oscillations in time.
We stress that these configurations differ from the so-called anomalous $\varphi_0$-junctions~\cite{Buz08,Gua20,Str20}, in which the ground state undergoes a finite phase shift, $\phi_0$, and an anomalous supercurrent can flow even at a zero phase bias.
In conclusion, apart from the relevance with respect to foundation aspects of the Josephson effects, frustrated Josephson systems can be used to achieve an arbitrary phase shift, rather than just 0 or $\pi$, towards on-chip phase batteries for biasing classical and quantum circuits, or for the design of superconducting memory and qubits.

\section{Conclusions} 
\label{Sec03}

We have demonstrated that in Josephson systems marked by multiple components with non-trivial phase frustration, a changeover of the ground state via non-harmonic drives generally yields an oscillating phase dynamics. The occurrence of this dynamical behavior is independent of the character of the transition, being observable for either continuous or abrupt variations of Josephson phases. The mechanism behind this finding can be ascribed to the intrinsic presence of discontinuous phase gradients in time across the transitions that cannot be avoided and naturally leads to the activation of dynamics. A key ingredient for generating the phase dynamics is the phase frustration of Josephson couplings and the consequent non-trivial phase configurations with values different from $0$ or $\pi$. Hence, in a scenario with multiple Josephson components we find that the rearrangement of Josephson phases across a transition among different ground states will always be accompanied by the activation of phase oscillations. This dynamics in turn has a time, and thus a frequency, behavior which is peculiar of the type of transition that the system undergoes. Thus, we argue that the activation of phase dynamics through non-harmonic external drives applied to a superconducting system is a clear-cut evidence of the presence of Josephson phase frustration or competing $0$ and $\pi$ channels. Moreover, the spectral character of the dynamics can be exploited to unveil the character of the transitions that are induced along the phase space trajectories.

Finally, although in a completely different context, we argue that the results obtained can be also applied to other physical cases where frustration plays an important role~\cite{Ram94}. For instance, the expression of Josephson energy in Eq.~\eqref{EqA01} is analogous to that of interacting spins with planar anisotropy and Heisenberg exchange that can be ferromagnetic ($0$-type Josephson coupling) or antiferromagnetic ($\pi$-type Josephson coupling). The scheme in Fig.~\ref{Fig01}(a) can indeed represent a system of interacting spins in a tetrahedral geometry.
By means of this analogy, we can thus predict that in a frustrated spin system with zero or non-vanishing net magnetization, the drive of a transition by varying the magnetic exchanges will always turn into a spin dynamics with generation of coherent or incoherent spin excitations associated to a change of orientation of the spin moments. 

Finally, we observe that the coupling between a multiband and an \emph{s}-wave superconductor can be in principle realized by sandwiching superconductors that are intrinsically multiband, as for the case of superconducting leads made by an iron-based and a conventional superconductor~\cite{Kal21,Ste21,Tia20,Has19,Kal18,Sch17,Dor15}. On the other hand, this system can be ad-hoc ``engineered'', as we proposed in Fig.~\ref{Fig01}(c) by exploiting a magnetic flux control, or alternatively through a setup designed by including additional ferromagnetic layers, which can provide $\pi$-pairing if inserted between the insulator and the superconductor~\cite{Gin16,Mar16,Wei06}. In this case, the temperature and thickness of the insulating and ferromagnetic layer serve as a control knob for tuning the Josephson couplings~\cite{Ban09,Kon02}. Finally, we mention the intriguing possibility of testing our theoretical predictions through multi-terminal JJs, which represent a research front that is currently yielding very interesting results~\cite{Riw16,Dra19,Pan20,Arn21,Arn22}.

In order to experimentally probe the time evolution of phase differences, the best strategy, especially when passing through a transition, is to look at the voltages, i.e. the phase velocity. A fast time-dependent response can be studied using Shapiro-like measurements through a microwave setup for Josephson emission~\cite{Boc18}. Then, one could look at the emission spectrum and see if it exhibits any characteristic peculiar of a transition~\cite{Kwo04}.

\begin{acknowledgments}
\indent This project has received funding from the European Union’s Horizon 2020 research and innovation programme under the Marie Sklodowska-Curie grant agreement No 841894. M.C., M.T.M and F.G. acknowledge support by the EU’s Horizon 2020 research and innovation program under Grant Agreement nr. 964398 (SUPERGATE). F.G. acknowledges the European Research Council under Grant Agreement No. 899315-TERASEC, and the EU’s Horizon 2020 research and innovation program under Grant Agreement No. 800923 (SUPERTED) for partial financial support.
\end{acknowledgments}

\appendix

\section{The ground states}
\label{AppA}

In the absence of magnetic field e current bias, the total energy of the system includes three interband and three interjuction contributions, see Eq.~\eqref{EqA01}.
By minimizing this equation with respect to $\theta_s$, $\theta_i$, and $\theta_j$ one can obtain the vector $\psi_{(n)}=(\phi_i,\phi_j,\phi_k)_{(n)}$ representing the ground state configurations of the system. In particular, by assuming $J_{is}=J_{js}=J_{ks}=J_{s}>0$ and $J_{ik}=J_{jk}=J$, the ground state results from the solution of the following system of equations
\begin{eqnarray}\label{EqA02}
&&J_{s}\sin\phi_i+J_{s}\sin\phi_j+J_{s}\sin\phi_k=0\\\nonumber
&&J_{s}\sin\phi_i+J_{ij}\sin\left ( \phi_i-\phi_j \right )+J\sin\left ( \phi_i-\phi_k \right )=0\\\nonumber
&&J_{s}\sin\phi_j-J_{ij}\sin\left ( \phi_i-\phi_j \right )+J\sin\left ( \phi_j-\phi_k \right )=0.\qquad
\end{eqnarray}

The Hessian matrix $\mathcal{H}$, which elements are obtained as $\mathcal{H}_{ij}=\frac{\partial^2 E}{\partial \phi_i \partial \phi_j}$, reads
\begin{widetext}
\[
\begin{scriptsize}
\mathcal{H}\!=\!\!\begin{bmatrix}\label{EqA03}
J_s\cos\phi_i+J_{ij}\cos\left ( \phi_i-\phi_j \right )+J\cos\left ( \phi_i-\phi_k \right ) & -J_{ij}\cos\left ( \phi_i-\phi_j \right ) & -J\cos\left ( \phi_i-\phi_k \right )\\ 
-J_{ij}\cos\left ( \phi_i-\phi_j \right ) & J_s\cos\phi_j+J_{ij}\cos\left ( \phi_i-\phi_j \right )+J\cos\left ( \phi_j-\phi_k \right ) & -J\cos\left ( \phi_j-\phi_k \right )\\ 
-J\cos\left ( \phi_i-\phi_k \right ) & -J\cos\left ( \phi_j-\phi_k \right ) & J_s\cos\phi_k+J\cos\left ( \phi_i-\phi_j \right )+J\cos\left ( \phi_i-\phi_k \right )\\
\end{bmatrix}
\end{scriptsize}
\]
\end{widetext}
The matrix $\mathcal{H}$ is symmetric and with off-diagonal terms.
For a given state to be stable, all the eigenvalues $\lambda_\mathcal{H}$ of the Hessian matrix $\mathcal{H}$ must be positive, i.e., the sum of the signs has to be equal to $\Sigma\text{sgn}(\lambda_\mathcal{H})=+3$.

Having assumed $J_{is}=J_{js}=J_{ks}=J_{s}>0$ and $J_{ik}=J_{jk}=J$, the ground states of the system can be expressed in a quite compact form. In particular, we obtain fourteen different solutions of the system of equations~\eqref{EqA02} that can be further grouped in three classes, see Fig.~\ref{Fig02}(a), labeled as ``trivial'', if given only by combinations of $0$ and/or $\pi$, and ``non-trivial'' [see Eqs.~\eqref{EqA04}-\eqref{EqA05}]. These solutions give quite different values of the total phase $\phi_{tot}=\phi_i+\phi_j+\phi_k$.

\begin{figure*}[t!!]
\includegraphics[width=2\columnwidth]{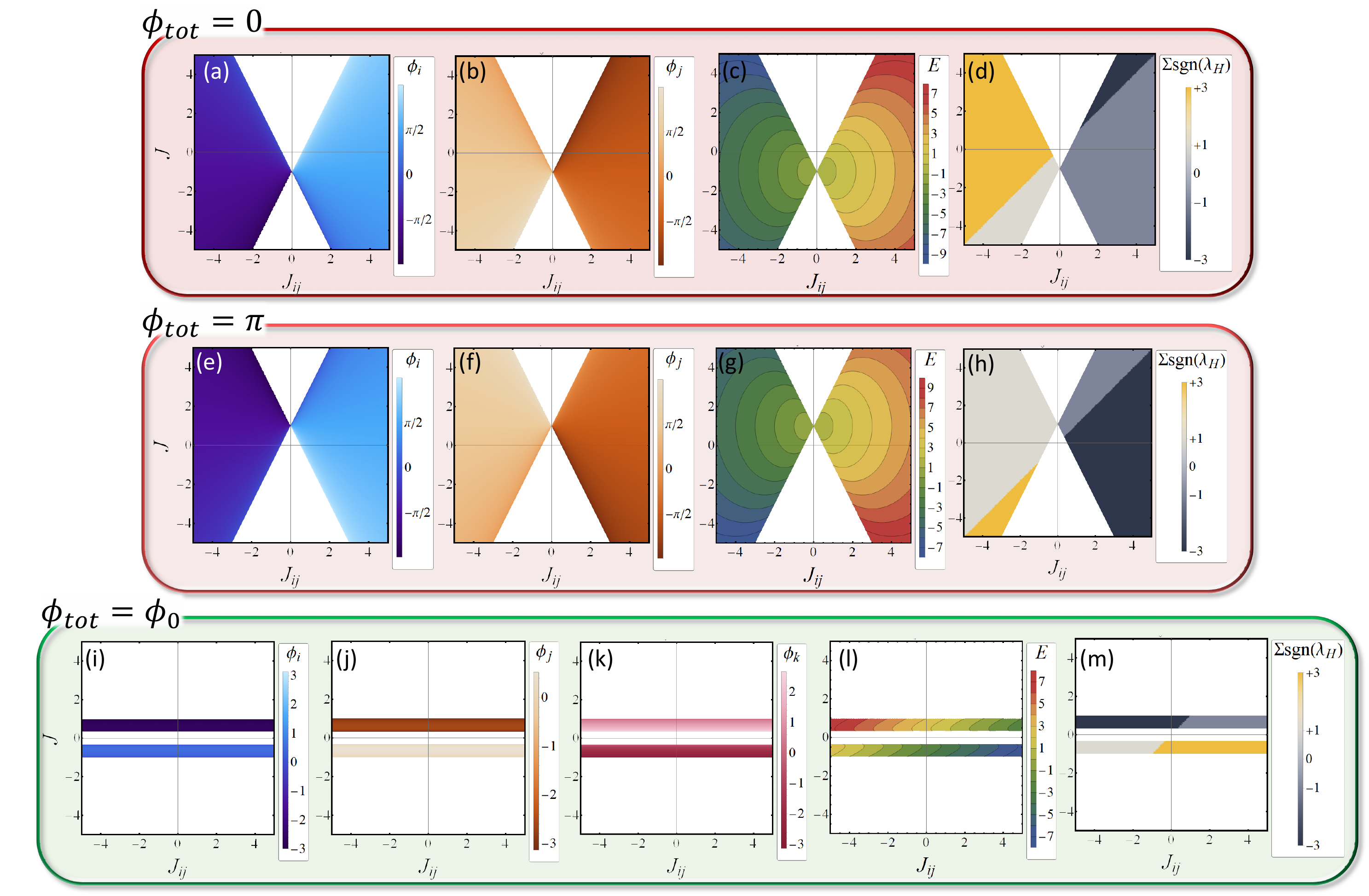}
\caption{Solutions $\psi_{(n)}=(\phi_i,\phi_j,\phi_k)_{(n)}$ of the system of equations~\eqref{EqA02}, total energy, $E$, and sum of the signs of the Hessian matrix eigenvalues, $\Sigma\text{sgn}(\lambda_\mathcal{H})$, as a function of $J$ and $J_{ij}$, in the non-trivial cases with $\phi_{tot}=0$ [top panels (a)-(d)],
$\pi$ [middle panels (e)-(h)], and $\phi_0$ [bottom panels (i)-(m)].
The white areas of the graphs represent $(J,J_{ij})$ combinations at which the system does not admit as possible solution the fundamental state under consideration.}
\label{FigSM-04}
\end{figure*}

In Fig.~\ref{FigSM-04}, we display the $(J,J_{ij})$-parameter space (here, the simplified notation in which $J\equiv J/J_s$ and $J_{ij}\equiv J_{ij}/J_s$ is used) of the solutions $\psi_{(n)}=(\phi_i,\phi_j,\phi_k)_{(n)}$, the total energy, $E$, and the sum of the signs of the Hessian matrix eigenvalues, $\Sigma\text{sgn}(\lambda_\mathcal{H})$, in the non-trivial cases with $\phi_{tot}=0$ [top panels (a)-(d)],
$\pi$ [middle panels (e)-(h)], and $\phi_0$ [bottom panels (i)-(m)].
The white areas of the graphs represent the combinations of the $(J,J_{ij})$ parameters for which the system does not admit as possible solution the $\psi_{(n)}$ ground state under consideration.

As previously noted, the total phase can even assume values different from $0$ and $\pi$, in which case, $\phi_{tot}=\phi_0^{\pm}\in[0,\pm\pi]$ depends only on the $J$ coupling according to
\begin{eqnarray}
\phi_0^{\pm}=&-&\arctan\left[3 J-\frac{1}{J},\pm\frac{\sqrt{J^2-(3J^2-1)^2}}{J}\right]\\\nonumber
&\pm&\arctan\left[-3 J-\frac{1}{J},\frac{\sqrt{J^2-(3J^2-1)^2}}{J}\right]\\\nonumber
&\mp&\arctan\left[-3 J-\frac{1}{J},-\frac{\sqrt{J^2-(3J^2-1)^2}}{J}\right].
\end{eqnarray}

We observe that the $\phi_0^{\pm}$ values tend to $\pm\pi$ for $J\to1$, while both converge to $0$ for $J\to-1/3$, see Fig.~\ref{Fig05}.\\

\section{The time-dependent model}
\label{AppB}

The equation of motion for the gauge-invariant phase differences can be derived from a Lagrangian approach taking a cue from Ref.~\onlinecite{Lin12}. 
The total Lagrangian of the system can be written as the sum of three contributions
\begin{equation}\label{EqB03}
\mathcal{L}=\mathcal{L}_1+\mathcal{L}_3+\mathcal{L}_{B}
\end{equation}
where the Lagrangian of the single- and the three-band superconductors are
\begin{widetext}
\begin{eqnarray}
\mathcal{L}_{1}=&&\frac{d}{8\pi \mu_{s}^{2}}\left [ A_{0}^{B}(r)+\frac{\Phi_0}{2\pi c}\partial_t\theta_s\left ( r,t \right ) \right ]^{2}-\frac{d}{8\pi \lambda_{s}^{2}}\left [ A_{x}^{B}(r)+\frac{\Phi_0}{2\pi c}\nabla\theta_s\left ( r,t \right ) \right ]^{2}	\\
\mathcal{L}_{3}=&& \sum_{z=i,j,k}\left \{ \frac{d}{8\pi\mu_{z}^{2}}\left [ A_{0}^{T}(r)+\frac{\Phi_0}{2\pi c}\partial_t\theta_z\left ( r,t \right ) \right ]^{2}
-\frac{d}{8\pi\lambda_{z}^{2}}\left [ A_{x}^{T}(r)+\frac{\Phi_0}{2\pi c}\nabla\theta_z\left ( r,t \right ) \right ]^{2}\right \}\\\nonumber
&&+\frac{\Phi_0}{2\pi c}\bigg [ J_{ij}\cos\left ( \theta_i-\theta_j \right )+J_{jk}\cos\left ( \theta_j-\theta_k \right )+J_{ik}\cos\left ( \theta_i-\theta_k \right ) \bigg ].
\end{eqnarray}
\end{widetext}

Here, $\mu_{s}$ and $\mu_{i}$ are the Thomas-Fermi lengths associated with charge screening, and $\lambda_{s}$ and $\lambda_{i}$ are the penetration depths for each band, $A_{0}^{B}$ ($A_{x}^{B}$) and $A_{0}^{T}$ ($A_{x}^{T}$) are electric (vector) potentials at the bottom and top electrodes, respectively, $d$ is the thickness of superconducting electrodes, and $\Phi_0=hc/2e$ is the magnetic flux quantum. 
The Lagrangian for the insulating barrier is
\begin{equation}\label{EqB03}
\mathcal{L}_{B}=\frac{b\epsilon_d}{8\pi}E_{b,z}^{2}-\frac{b}{8\pi}B_{b,z}^{2}-V_J
\end{equation}
where $b$ is the thickness of the barrier and $\epsilon_b$ the dielectric constant. The electric and magnetic fields in the barrier are
\begin{equation}\label{EqB04}
E_{b,z}=-\frac{1}{c}\partial_t A_{b,z}-\partial_z A_0=-\frac{1}{c}\partial_t A_{b,z}-\frac{A_0^T-A_0^B}{b}
\end{equation}
and
\begin{equation}\label{EqB05}
B_{b,y}=\partial_z A_x-\partial_x A_{b,z}=\frac{A_x^T-A_x^B}{b}-\partial_x A_{b,z},
\end{equation}
while the Josephson coupling $V_J$ is
\begin{equation}\label{EqB06}
V_J=-\frac{\Phi_0}{2\pi c}\sum_{z=i,j,k}J_{zs}\cos\left ( \phi_{zs} \right )
\end{equation}
withe the gauge-invariant phase difference
\begin{equation}\label{EqB07}
\phi_{zs}=\theta_z-\theta_s-\frac{2\pi b}{\Phi_0} A_{b,z}.
\end{equation}

%

%
%
At a temperature well below the critical value, according to the microscopic theory~\cite{Bar82,Agt02} the Josephson couplings can be written as
\begin{equation}\label{EqB07a}
J_{si}=\frac{2\hbar}{eR_{bi}}\frac{\left|\Delta_i\Delta_s \right|}{\left|\Delta_i \right|+\left|\Delta_s \right|}K\left ( \frac{\left|\Delta_i \right|-\left|\Delta_s \right|}{\left|\Delta_i \right|+\left|\Delta_s \right|} \right ),
\end{equation}
where $\Delta_i$ and $\Delta_s$ are the superconducting gap of different condensates, $K(x)$ is the complete elliptic integral of the first kind, and $R_{bi}=\hbar^3\big/\left ( 4\pi e^2N_i(0)N_s(0)t_{i,s} \right )$ is the resistance for the $i$-th channel, with $N_i(0)$ and $N_s(0)$ being the density of states of quasiparticles in the $i$-th band and the \emph{s}-wave superconductor, respectively, and $t_{i,s}$ the tunneling probability for electrons between two superconductors. 

For the sake of convenience, we normalized the space to $\lambda_{c_1}=\sqrt{c\Phi_0/(8\pi^2bJ_{s1})}$ and the time to the inverse of $\omega_{c_1}=c/\left ( \sqrt{\epsilon_d}\lambda_{c_1} \right )$, while the magnetic and electric field are written in units of $\Phi_0/(2\pi \lambda_{c_1}b)$ and $\Phi_0\omega_{p_1}/(2\pi cb)$, respectively.

The three equations of motion that we need can be obtained by applying the Euler-Lagrangian equation with respect to $A_{b,z}$, so to obtain the Ampere’s law
\begin{equation}\label{EqB08}
\partial_x B_{b,y}=\sum_{z=i,j,k}J_{zs}\sin\left ( \phi_{zs} \right )+\partial_t E_{b,y}.
\end{equation}
In the spatially independent case and assuming $J_{is}=J_{js}=J_{ks}=J_{s}>0$ and $J_{ik}=J_{jk}=J$, we obtain
\begin{equation}\label{EqB09}
\sum_{z=i,j,k}\left [ J_{s}\sin\left ( \phi_{zs} \right )+\frac{\partial_t^2 \phi_{zs}}{C_e \alpha_z} \right ]=0
\end{equation}
where $C_e=(1+\epsilon_d\,\alpha_s)\sum_{z}\frac{1}{\alpha_z}+\epsilon_d$ and $\alpha_{z(s)}=\mu_{z(s)}^2/(db)$.
The remaining two necessary equations for the gauge-invariant phase differences can be obtained by properly combining the equations deriving by variation of $\mathcal{L}$ with respect to $\theta_s$, $\theta_z$, with $z=i,j,k$. In this way, we obtain
\begin{widetext}
\begin{eqnarray}\label{EqB10}
&&\frac{\partial_t^2 \phi_{is}}{\epsilon_d\alpha_i}+\sum_{z=i,j,k}\left [ J_{s}\sin\left ( \phi_{zs} \right )+k_i\frac{\partial_t^2 \phi_{zs}}{\alpha_z} \right ]+\frac{\xi_i}{\xi_s}\left [ J_{s}\sin\left ( \phi_{is} \right ) +J_{ij}\sin\left ( \phi_{is}-\phi_{js} \right )+J\sin\left ( \phi_{is}-\phi_{ks} \right )\right ]=0\qquad\\\label{EqB11}
&&\frac{\partial_t^2 \phi_{js}}{\epsilon_d\alpha_j}+\sum_{z=i,j,k}\left [ J_{s}\sin\left ( \phi_{zs} \right )+k_j\frac{\partial_t^2 \phi_{zs}}{\alpha_z} \right ]+\frac{\xi_j}{\xi_s}\left [ J_{s}\sin\left ( \phi_{js} \right ) -J_{ij}\sin\left ( \phi_{is}-\phi_{js} \right )+J\sin\left ( \phi_{js}-\phi_{ks} \right )\right ]=0\qquad
\end{eqnarray}
\end{widetext}

where $k_z=\frac{1}{C_e}\left ( 1-\frac{1}{\epsilon_d \alpha_z}-\frac{\alpha_s}{ \alpha_z} \right )$ and $\xi_{z(s)}=\lambda_{z(s)}^2/(db)$.

The phase dynamics of the junction is described by Eqs.~\eqref{EqB09}-\eqref{EqB11}.

To compute the time evolution of the phases, we take for convenience of calculation $\alpha_s=\alpha_i=\alpha_j=\alpha_k=0.1$ and $\xi_s=\xi_i=\xi_j=\xi_k=\xi$.

Equation~\eqref{EqB09} reveals that the capacitive terms in the differential equations are proportional to the coefficient $1/(\alpha C_e)$, which tends to increases when $\alpha$ is reduced. Moreover, we observe that the parameter $\alpha$ is inversely proportional to the barrier thickness. Thus, we expect the establishment of a predominant overdamped regime as we reduce $\alpha$, i.e, as we increase the barrier thickness. We therefore trust that the phenomenology described in this work will remain qualitatively unchanged, but that the specificities of the dynamics may depend on the choice of $\alpha$, whose value, in particular, may result in an under- or over-damped dynamic regime.


\section{Other transitions}
\label{AppC}

\begin{figure*}[t!!]
\includegraphics[width=2\columnwidth]{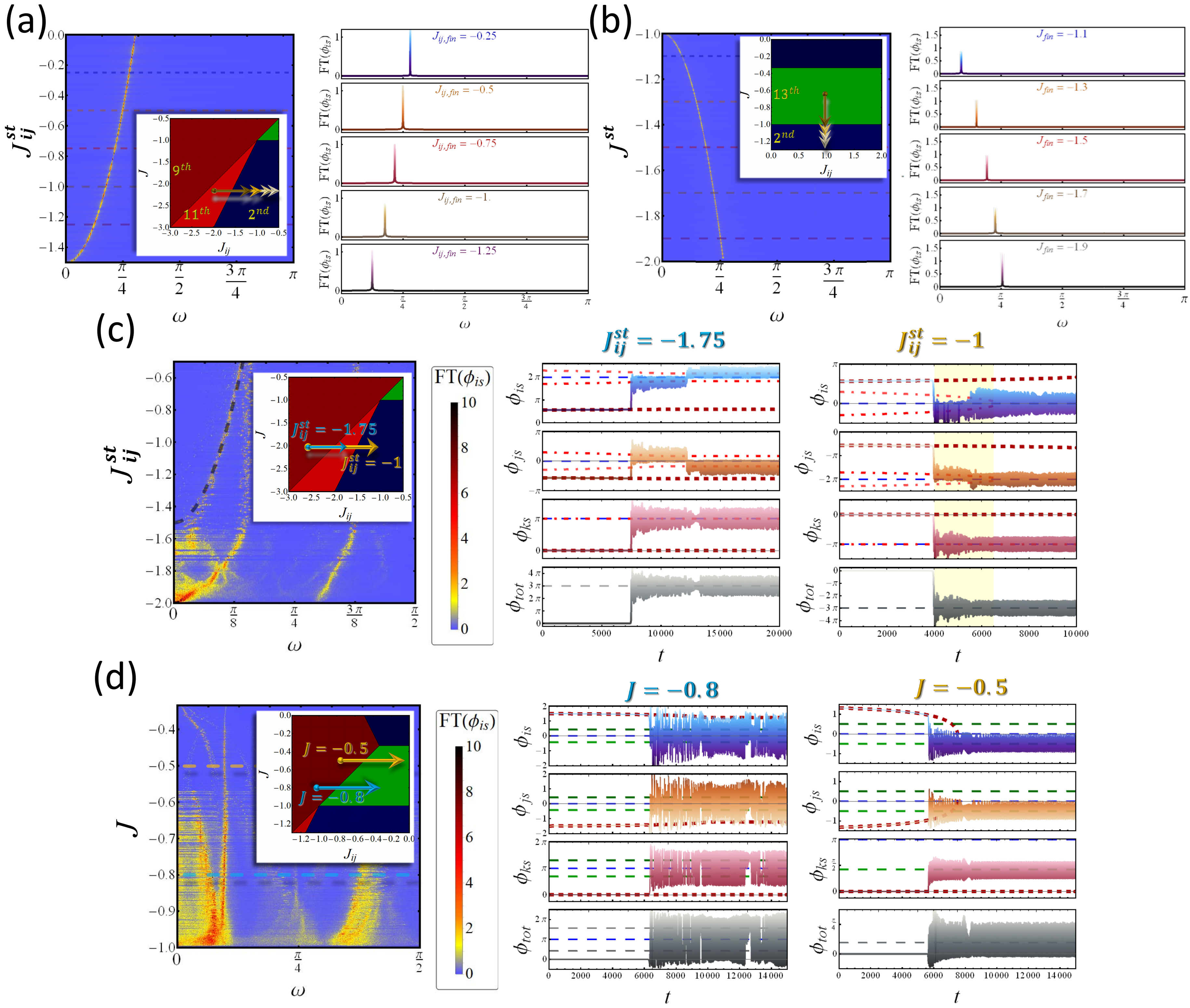}
\caption{(a) FT of $\phi_i(t)$ as a function of $J_{ij}^{st}\in[-1.5,0]$, with $J=-2$ and $J_{ij}(t)\in[-2,J_{ij}^{st}]$ and some selected profiles. 
(b) FT of $\phi_i(t)$ as a function of $J^{st}\in[-2,-1]$, with $J_{ij}=-1$ and $J(t)\in[-0.675,J^{st}]$ and some selected profiles.
(c) FT of $\phi_i(t)$ as a function of $J_{ij}^{st}\in[-2,0]$, with $J=-2$ and $J_{ij}(t)\in[-2.5,J_{ij}^{st}]$ and phase evolutions at two values of $J_{ij}^{st}=-1.75$ and $-1$, corresponding to incoherent and coherent frequency responses, respectively.
(d) FT of $\phi_i(t)$ as a function of $J\in[-1,-0.33]$, with $J_{ij}(t)\in[J-0.5,J+0.5]$ and phase evolutions at two values of $J=-0.8$ and $-0.5$, corresponding to incoherent and coherent frequency responses, respectively. Each density plot contains also an inset showing the $(J,J_{ij})$-phase diagram of the lowest-energy GS and some arrows highlighting the phase transitions on which we focus. The blue, red, and green dashed lines in the phase dynamics plots indicate the analytical solutions listed in Fig.~\ref{Fig02}(a) around which the phases evolve.}
\label{Fig07}
\end{figure*}

In this appendix we shed light on the other transitions that can take place in this system.

First, we underline that the choice of the GS $\psi_{(9)}$ as initial state, i.e., as in the cases discussed in the main text, was dictated mainly by the fact that starting from this, by changing only the value of one coupling and always using the same kind of drive it was possible to explore all possible interesting transitions. Specifically, this means the feasibility to switch from a non-trivial to a trivial state leaving the value of $\phi_{tot}$ unchanged [e.g., Figs.~\ref{Fig03}(a) and~\ref{Fig04}(a)] and from a non-trivial to another non-trivial state, with $\phi_{tot}$ changing of $\pi$ [e.g., Figs.~\ref{Fig03}(b) and~\ref{Fig04}(b)] or $\phi_{tot}\in(0,\pm\pi)$ [e.g., Figs.~\ref{Fig03}(c) and~\ref{Fig04}(c)]. 

In Fig.~\ref{Fig07}(a) and (b) we demonstrate other two possibilities. In particular, referring to the parameter space in Fig.~\ref{Fig02}b, we present the $\pi\to\pi$ transition ``from light-red to blue'' [see Fig.~\ref{Fig07}(a) obtained starting from the non-trivial GSs $\psi_{(11)}$] and the $\phi_0\to0$ transitions ``from green to blue'' [see Fig.~\ref{Fig07}(b) obtained starting from the non-trivial GSs $\psi_{(13)}$]. In both cases, the FT of $\phi_i(t)$ is highly coherent, as the selected FT profiles shown in these panels well demonstrate. 

Figure~\ref{Fig07} also helps to better understand the origin of the coherent FT response evinced in Fig.~\ref{Fig04}(b) and (c). 

Figure~\ref{Fig04}(b) shows the FT of $\phi_{tot}(t)$ as a function of $J_{ij}^{st}$, and demonstrates that for $J_{ij}^{st}\gtrsim-1.5$ the frequency response is composed by two well-defined peaks, which intimately depend on the underlying phase dynamics. To understand their nature, we refer to Fig.~\ref{Fig07}(c), in which, for the sake of clearness, we present both the FT of $\phi_{i}(t)$ versus $J_{ij}^{st}$ and the phase evolutions at two different values of $J_{ij}^{st}$. 

The upper part of the density plot in FT of $\phi_{i}(t)$ in Fig.~\ref{Fig07}(c) shows two kind of peaks, one dependent on $J_{ij}^{st}$, which is marked by a black dashed line, and one independent of $J_{ij}^{st}$. The former, is of the same nature as the FT peak shown in Fig.~\ref{Fig04}(a) [please note that this peak is not present in Fig.~\ref{Fig04}(b) since this density plot shows the FT of $\phi_{tot}(t)$]. On the contrary, the latter comes from the fact that the system remembers being ``passed'' through non-trivial GSs $\psi_{(11,12)}$, even if it definitively ends up in the trivial state $\psi_{(1)}$. Thus, these peaks are ``frozen'' to the last characteristic frequency of the nontrivial GS through which the system passed, and this is why they do not change anymore by increasing $J_{ij}^{st}$ further. 
This phase dynamics is demonstrated in the middle panel of Fig.~\ref{Fig07}(c), obtained by ending the drives at $J_{ij}^{st}=-1.75$; in this case, the phases eventually resides in the two non-trivial GSs $\psi_{(11,12)}$ after the linear drive is switched off. Instead, in the right panel of Fig.~\ref{Fig07}(c) the drive stops at $J_{ij}^{st}=-1$, so that the system resides in the trivial GS $\psi_{(1)}$ after staying for a while in the non-trivial GSs $\psi_{(11)}$ (see the yellow shaded region). 
In conclusion, for the $\pi\to\pi$ transition the two FT peaks are not merely commensurate, but reflect the possibility that the phases evolve through different GSs.

Finally, we note that in Fig.~\ref{Fig03}(c) the system ``populates'' both the GSs $\psi_{(13,14)}$. This occurs since they are indeed ``very close'', i.e., the energy barrier between them is small. This observation helps to understand the incoherent/coherent response observed in the $0\to\phi_0$ transitions as $J_{ij}^{st}$ is changed, see Fig.~\ref{Fig04}(c). In fact, the incoherent frequency response shown for $J_{ij}^{st}\to-1$ in Fig.~\ref{Fig04}(c) is given by the fact that the two $\psi_{(13,14)}$ potentials minima are quite close to the trivial $\psi_{(7)}=(0,0,-\pi)$ GS, so that the phases evolve through all these three states, see the middle panel of Fig.~\ref{Fig07}(d) for $J_{ij}^{st}=-0.8$.
Instead, when $J_{ij}^{st}$ is increased, i.e., $J_{ij}^{st}\to-1/3$, the system ``chooses'' just one of the two non-trivial GSs $\psi_{(13,14)}$, see the right panel of Fig.~\ref{Fig07}(d) for $J_{ij}^{st}=0.5$. This gives the observed coherent frequency response shown in Figs.~\ref{Fig04}(c) and (f), with the characteristic frequency in this case becoming the plasma mode in this specific potential well.


%

\end{document}